\documentclass[twocolumn,english,aps]{revtex4}
\usepackage[latin1]{inputenc}
\usepackage{graphicx}

\makeatletter

\makeatletter

\usepackage{babel}
\makeatother
\begin{document}

\title{Isochronal synchronization of delay-coupled systems}

\author{Ira B.~Schwartz and Leah B. Shaw}

\affiliation{US Naval Research Laboratory, Code 6792, Nonlinear Systems Dynamics
Section, Plasma Physics Division, Washington, DC 20375}

\begin{abstract}
We consider small network models for mutually delay-coupled systems
which typically do not exhibit stable isochronally synchronized
solutions.  We show analytically and numerically that for
certain coupling architectures which involve delayed self feedback to
the nodes, the oscillators become isochronally
synchronized. Applications are shown for both incoherent pump coupled
lasers and spatio-temporal coupled fiber ring lasers.
\end{abstract}
\maketitle

\section{Introduction}

Synchronization of networked, or coupled, systems has been
examined for large networks of identical  \cite{pikovsky}
and  heterogeneous oscillators \cite{RestrepoOH06}. For
coupled systems with smaller numbers of oscillators, several new dynamical phenomena
have been observed, including generalized \cite{RulkovSTA95}, phase \cite{RosenblumPK96},
and lag \cite{RosenblumPK97} synchronization. Lag synchronization,
in which there is a phase shift between observed signals, is one of
the routes to complete synchrony as coupling is increased \cite{RosenblumPK97} and
may occur without the presence of delay in the coupling terms.

For systems with delayed coupling, a time lag between the oscillators
is typically observed, with a leading time series followed by a lagging one.
Such lagged systems are said to exhibit achronal synchronization.
In \cite{HeilFEMM01}, the existence of achronal synchronization in
a mutually delay-coupled semiconductor laser system was shown
experimentally, and in \cite{WhiteMM02},  studied theoretically in a
single-mode semiconductor laser model. 
In the case of short coupling delay for unidirectionally coupled systems,
 anticipatory synchronization occurs when a response in a system's
state is not replicated simultaneously but instead is anticipated by the response
system \cite{Voss00,Voss01}, and  an example of anticipation in synchronization
is found in coupled semiconductor lasers \cite{Masoller01}. Cross-correlation
statistics between the two intensities showed clear maxima at delay
times consisting of the difference between the feedback and the coupling
delay. Anticipatory responses in the presence of stochastic effects
have been observed in models of excitable media \cite{CiszakCMMT03},
 as well as in experiments of coupled semi-conductor lasers in a
transmitter-receiver configuration \cite{Siva_PhysRevLett.87.154101}.
{\bf When the zero lag state is unstable and achronal synchronization 
occurs, the situation may be further complicated by switching between 
leader and follower.  Switching has been observed theoretically and 
experimentally in stochastic systems \cite{MuletMHF04} but may occur even 
in deterministic chaotic systems \cite{WuZ03}.}

Given that both lag and anticipatory dynamics may be observed in delay-coupled
systems, it is natural to ask whether the isochronal, or zero lag, state, in which
there is no phase difference in the synchronized time series, may
be stabilized in coupled systems. {\bf Stabilizing the isochronal state is important
in bi-directional chaotic communication
systems, as shown in  recent theoretical  work on communicating in systems with delay
\cite{zHOU07_citeulike:1107329}.} Stable
isochronal synchronization of semiconductor lasers has been observed 
recently in experiments \cite{Tangetal04,Klein06} and numerically
\cite{VicenteTMML06,Klein06}.  {\bf Examples of partial
isochronal synchrony in which only some of the oscillators in a delay
network synchronize may be found in 
\cite{Fischer06citeulike:1107314,zHOU07_citeulike:1107329}, 
and recently a
theoretical explanation for partial synchronization has appeared in
\cite{Landsman07_citeulike:1086005}.} Other examples of isochronal
synchrony have appeared in  neural network models with delay 
\cite{DhamalaJD04,RossoniCDF05}.

In this letter, we explore the possibility
of adding self feedback to two globally coupled situations: 1.
Incoherent delay-coupled semiconductor systems \cite{KimRACS05},
and 2. Coupled spatio-temporal systems consisting of coupled
fiber ring lasers \cite{WilliamsGR97} with delay \cite{ShawSRR06}. 

We consider $N$ coupled oscillators of the following form. Let $F$
denote an $m$-dimensional vector field, $B$ an $m\times m$ matrix, and $\kappa_{j}$, where $j=1\cdots N$, denote
the coupling constants. For the cases we examine here, we consider
global coupling including self feedback:
\begin{equation}
\frac{dx_{i}(t)}{dt}=F(x_{i}(t),x_{i}(t-\tau))+\begin{array}{c}
\\\sum\\
j\ne i\end{array}\kappa_{j}Bx_{j}(t-\tau).\label{eq:GeneralSystem}\end{equation}

Given the structure of Eq.~\ref{eq:GeneralSystem}, we examine the
stability transverse to the synchronized state, $S=\{x_i(t): x_{i}(t)=s(t),i=1\cdots N\}$, by
defining $\eta_{ij}\equiv x_{j}-x_{i}.$ The linearized variations in the
direction transverse to $S$ are then given by 
\begin{equation}
\begin{array}{c}
\frac{d\eta_{ij}(t)}{dt}=D_{1}F(x_{i}(t),x_{i}(t-\tau))\eta_{ij}(t)\\
+D_{2}F(x_{i}(t),x_{i}(t-\tau))\eta_{ij}(t-\tau)\\
+(\kappa_{i}-\kappa_{j})Bx_{i}(t-\tau)-\kappa_{j}B\eta_{ij}(t-\tau)\end{array}\label{eq:LinVarGenSysa}\end{equation}
where $D_{i}$ denotes the partial derivative with respect to the $i^{th}$ argument.

We make the following hypotheses to simplify the analysis: (H1): Assume that the dependence on the time delayed variables in $F$ takes the same form as the delay coupling; i.e., $D_{2}F(x,y)=B\kappa_{f}$. (H2): Let $\kappa_{i}=\kappa_{f}=\kappa,i=1\cdots N.$ Equation
\ref{eq:LinVarGenSysa} then simplifies to 
\begin{equation}
\frac{d\eta_{ij}(t)}{dt}=D_{1}F(x_{i}(t),x_{i}(t-\tau))\eta_{ij}(t),\label{eq:LinVar}\end{equation}
where it is understood the arguments of the derivatives are computed
along the synchronized solution $s(t)$, and the solution is a function of parameters such
as coupling and delay. Computing Eq.~\ref{eq:LinVar} along the synchronized
state will generate the Lyapunov exponents for the transverse directions,
and we examine the effect of coupling and delay
by computing the cross-correlations between time series as well.

To examine the stability  of the isochronally synchronized state of
Eq.~\ref{eq:GeneralSystem}, we model $N=3$  lasers
that are pump coupled \cite{KimRACS05,CarrTS06}. An isolated semiconductor
laser's dynamics at the $i^{th}$ node is governed by $\frac{dz_{i}}{dt}=\bar{F}(z_{i}),$$z_{i}=(x_{i},y_{i})$,
where
\begin{equation}
\bar{F}(z)=\left[-y-\epsilon x(a+by),x(1+y)\right],\label{eq:SO model}\end{equation}
and $x,y$ are the scaled carrier fluctuation number and normalized
intensity fluctuations about steady state zero, respectively. 
 $\epsilon^{2}$ is the ratio
of photon to carrier lifetimes, and $a$ and $b$ are dimensionless
constants (see \cite{SchwartzE94} for details on the derivation).

\begin{figure}
\includegraphics[width=2.55in,height=3.5in,keepaspectratio]{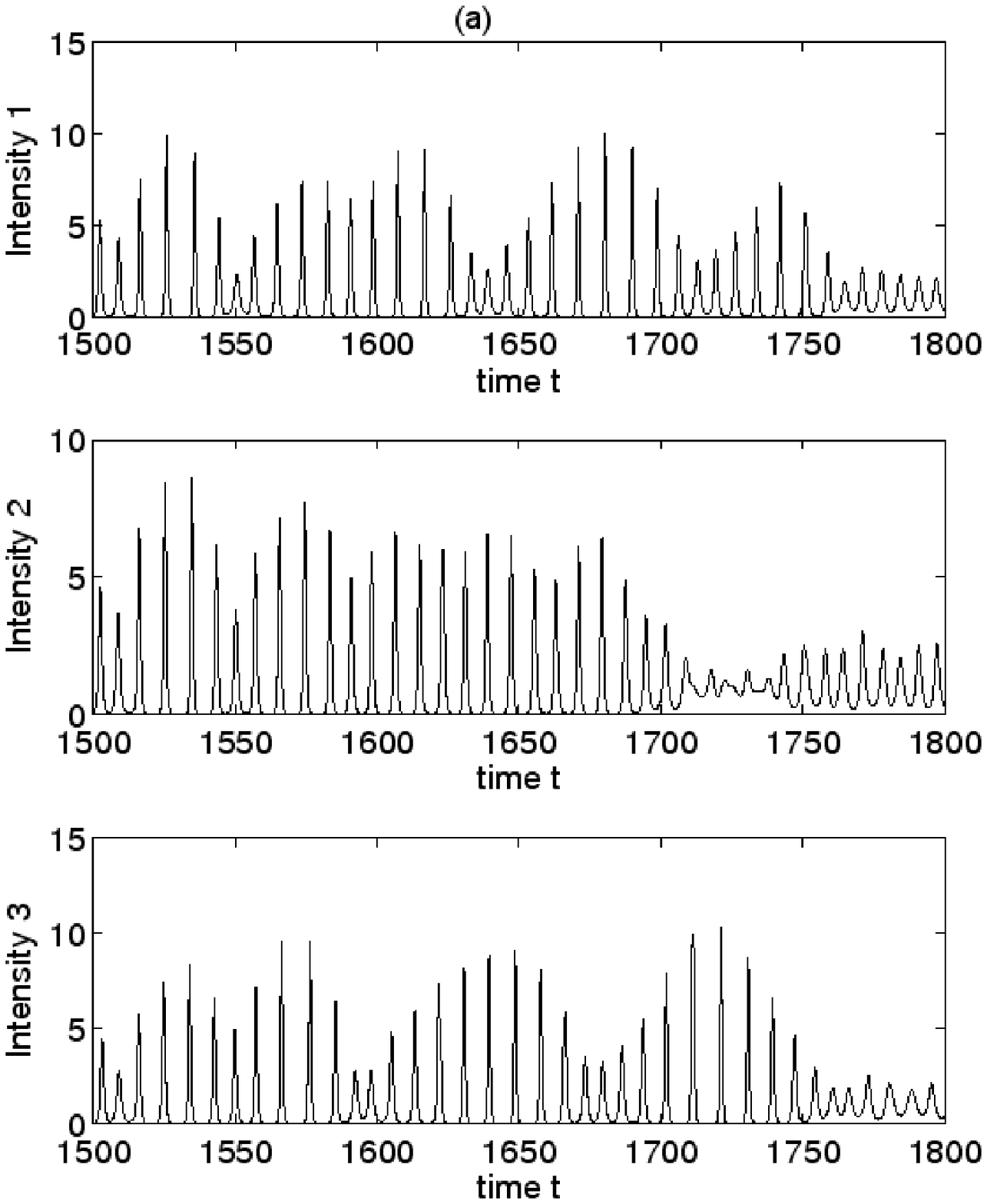}
\includegraphics[width=4.0in,height=3.5in,keepaspectratio]{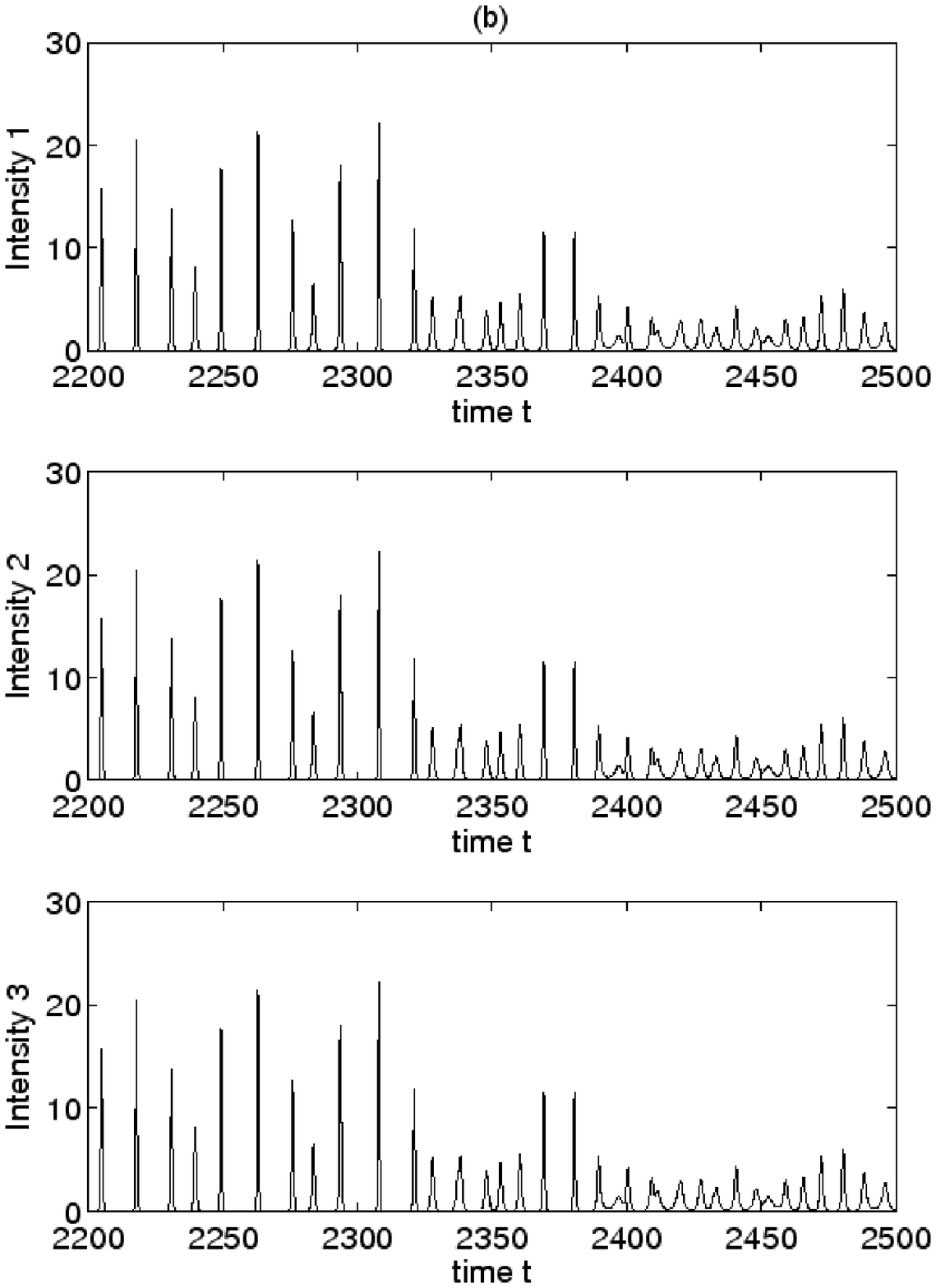}

\caption{\label{cap:TSdelayexample} An example of delay-coupled dynamics
showing intensities computed for $N=3,\kappa=3.0\epsilon,\tau=30,a=2,b=1,\epsilon=\sqrt{0.001}$, using
Eq.~\ref{eq:SO model}. (a) shows  a solution
where the lasers are coupled globally without self feedback, in which isochronal synchrony does not occur.
(b) shows a stable isochronal solution with self feedback terms
included.}
\end{figure}

The coupling strengths
are $\kappa_{i}=\kappa_{f}=\kappa,i=1,2,3$. This leads to the following
set of differential equations for the system: 
\begin{equation}
\frac{dz_{i}(t)}{dt}=\bar{F}(z_{i}(t))+\begin{array}{c}
\begin{array}{c}
3\\
\kappa\sum\\
i=1\end{array}Bz_{i}(t-\tau),i=1,2,3\end{array},\label{eq:three Oscillators}\end{equation}
where $m=2$, and $B(1,2)=1$, with all other entries in $B$ equal to $0$.
An example of the intensities  with and without self feedback  in
  Fig.~\ref{cap:TSdelayexample} shows explicitly the
  effect of self feedback in stabilizing the isochronal solution.
Writing down the differential equation for the transverse directions
in matrix form
for Eq.~\ref{eq:three Oscillators} using Eq.~\ref{eq:LinVar} and
expanding near the synchronized solution $\eta_{ij}=0,$ we obtain
$X^\prime (t)=A(t,\kappa,\tau,\epsilon)X(t)$, where $A(t,\kappa,\tau,\epsilon)=DF(s(t,\kappa,\tau,\epsilon))$, and $X(0)=I.$  
Due to the nature of the global coupling with self feedback, each node
receives the same signal. Therefore, the transverse stability does
not explicitly depend on the coupling or delay, but rather on the
dynamics of local nodes \cite{PecoraC98}. To examine the stability
of the isochronal state, we derive some
properties of the transverse Lyapunov exponents (TLE). The
TLE satisfy the following limit: $\lambda(x_{0},y_{0},u)=\lim_{t\rightarrow\infty}\frac{1}{t}\log\frac{\left\Vert X(t)u\right\Vert }{\left\Vert u\right\Vert }.$
Here $u$ is a vector in a given direction. 

By computing the solution to the linear variational equations along
a given solution, we can extract the TLE. To examine the
scaling behavior of the TLE, let $\Delta(t,\kappa,\tau,\epsilon)=\det(X(t,\kappa,\tau,\epsilon))$.
Then, we have that $\Delta(t,\kappa,\tau,\epsilon)=\exp({\displaystyle \int_{0}^{t}trace(A(s,\kappa,\tau,\epsilon))ds)}$
\cite{Hartman:1982:ODE}. Taking the log of the matrix solution, and noting the determinant
of a matrix is the product of its eigenvalues, we have:
\begin{eqnarray}
\sum_{i=1}^{m}\lambda(x_{0},y_{0},e_{i}) & = &
\lim_{t\rightarrow\infty}\frac{1}{t}\log|\det(X(t,\kappa,\tau,\epsilon)|,\label{eq:sum
  of exponents}\end{eqnarray}
where  $e_i$  are arbitrary independent basis vectors. 
Equation \ref{eq:sum of exponents} yields a rate of volume change
in the dynamics in the transverse directions. The solution may 
still be chaotic with one or more exponents being positive, but if
sufficiently dissipative, volumes will shrink over time.

\begin{figure}
\includegraphics[width=2.5in,height=2.5in,keepaspectratio]{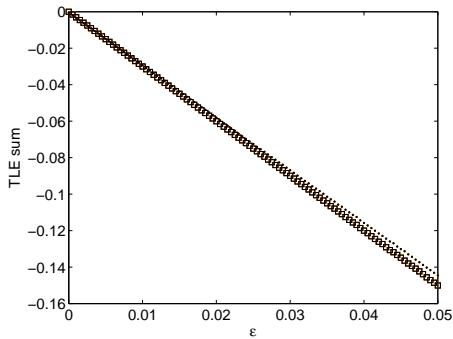}

\caption{\label{cap:predictingSumLE}  Prediction of the scaling of the sum of transverse
Lyapunov exponents for Eq.~\ref{eq:three Oscillators} with respect to $\epsilon$. Other parameter values are as in Figure \ref{cap:TSdelayexample}(b). Squares
are the prediction using Eq.~\ref{eq:LyapSum}, and dots are the numerical
values.}
\end{figure}

From Eq.~\ref{eq:SO model}, since $trace(A(t,\kappa,\tau,\epsilon))=-\epsilon(a+by(t,\kappa,\tau,\epsilon))+x(t,\kappa,\tau,\epsilon)$,
and assuming the inversion, $x(t,\kappa,\tau,\epsilon),$ has zero time average
due to symmetry (which is observed numerically \cite{LBIBS06}), we have $\int_{0}^{t}trace(A(s,\kappa,\tau,\epsilon))ds=-\epsilon(a+b\left<y_{\kappa,\tau,\epsilon}\right>)t$
and from Eq.~\ref{eq:sum of exponents}, we have \begin{equation}
\lambda(x_{0},y_{0},e_{1})+\lambda(x_{0},y_{0},e_{2})=-\epsilon(a+b\left<y_{\kappa,\tau,\epsilon}\right>).\label{eq:LyapSum}\end{equation}
Since $\epsilon$ appears explicitly, it is easy to see how the sum of the TLE
scales with $\epsilon$ and compares with numerical experiments as in Fig.
\ref{cap:predictingSumLE}. 

\begin{figure}
\includegraphics[width=3.5in,keepaspectratio]{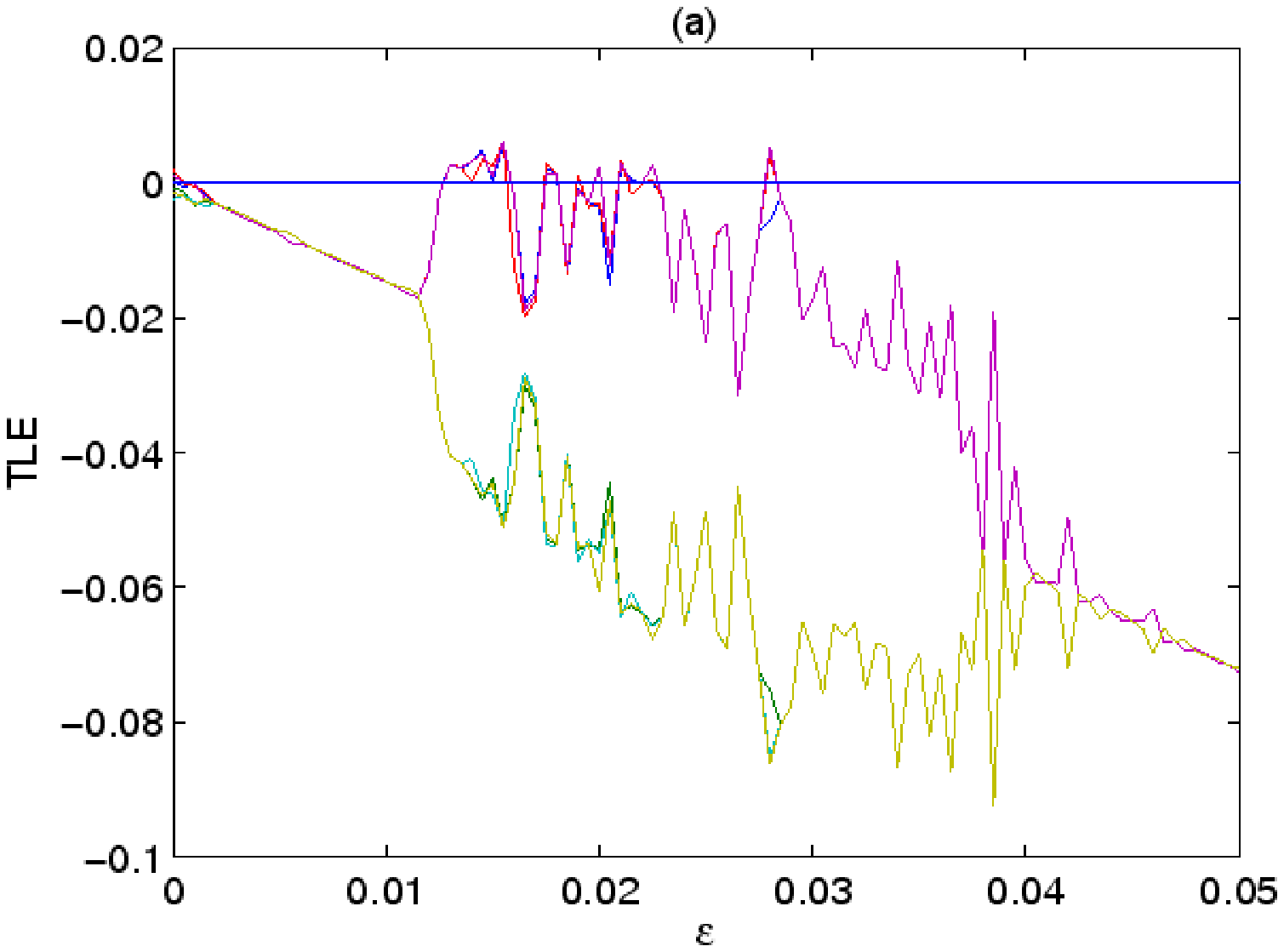}
\includegraphics[width=3.5in,keepaspectratio]{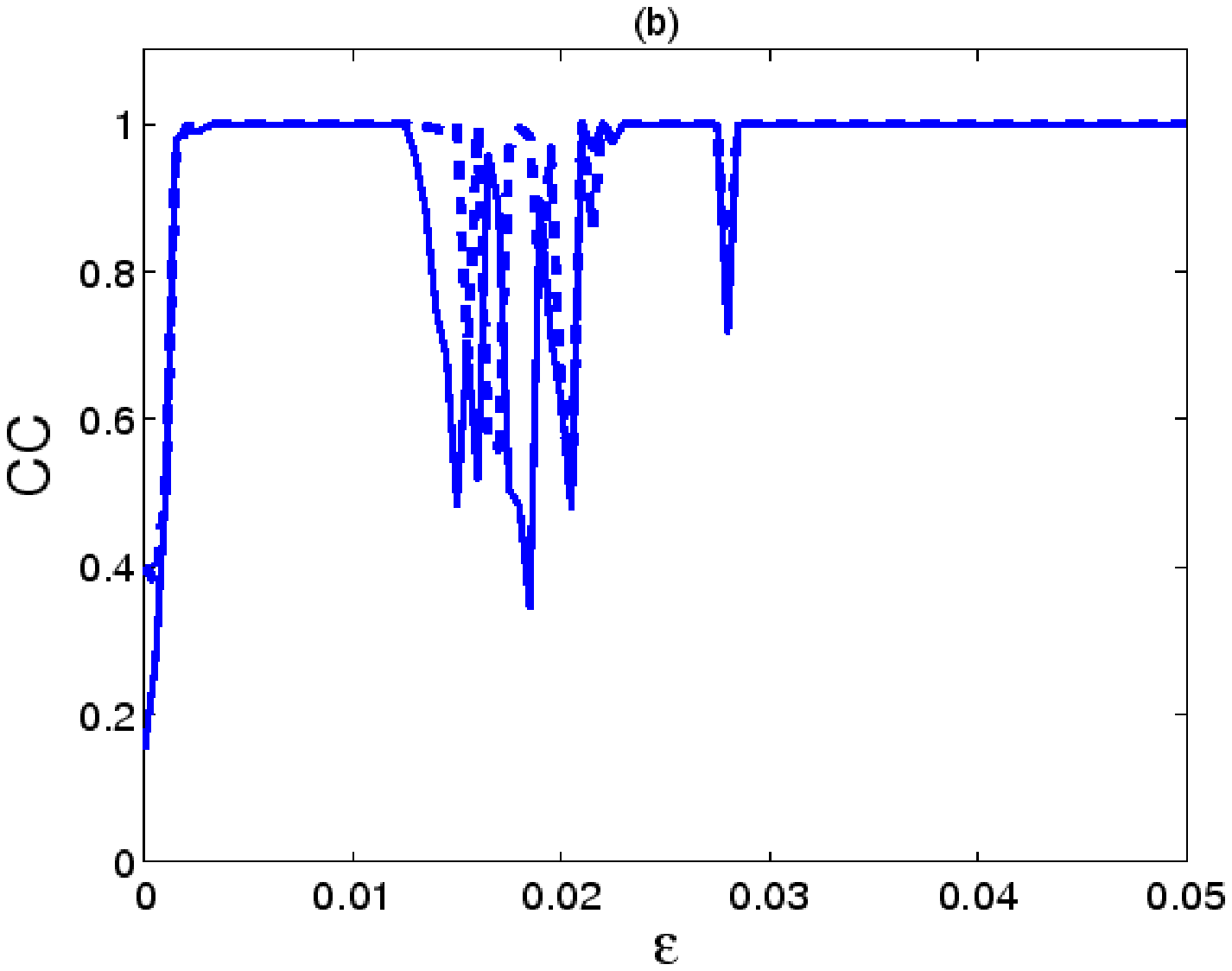}

\caption{\label{cap:CC} (Color online) (a) Transverse Lyapunov exponents  
and (b)
  cross-correlation (CC) of the dynamics for the same conditions as in 
Fig. \ref{cap:predictingSumLE}.  \textbf{In (b), the cross-correlations} 
between lasers 1 
and 2 (solid line) and 1 and 3 (dashed line) are shown.  For most values of $\epsilon$ shown here, a cross-correlation of 1 is achieved when the shift between the time traces is zero, showing that the isochronal solution is stable.}
\end{figure}

\begin{figure}
\includegraphics[width=3.5in,height=2.0in,keepaspectratio]{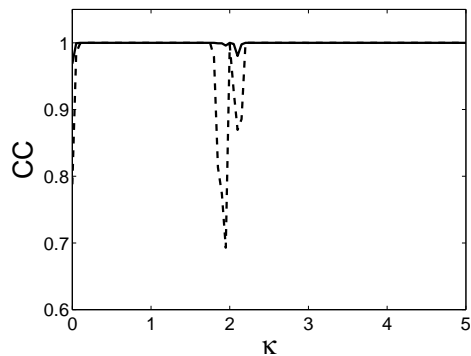}

\caption{\label{cap:CCLEvsdelay} Cross-correlation (CC) between lasers 1 and 2 (solid line) and between 1 and 3 (dashed line) vs.~coupling $\kappa$ for
  Eq.~\ref{eq:three Oscillators}.  Other parameters are the same as in Figure \ref{cap:TSdelayexample}(b).}
\end{figure}

Although the sum of the TLE is negative, loss of synchrony due to instability may occur at intermediate values of $\epsilon$, as seen in Fig.~\ref{cap:CC}.  Regions where the isochronally synchronized solution is unstable are associated with one or more positive transverse Lyapunov exponenents.  
On the other hand, for sufficiently large damping, the transverse
exponents reveal a stronger overall reduction in the phase space volume.
The stability of isochronal synchrony with respect to other parameters
can also be computed, e.g., as  shown in Fig.~\ref{cap:CCLEvsdelay} for variations in coupling strength $\kappa$.

To illustrate the robustness of the self feedback structure for generating isochronal synchronization in delay
coupled systems, we examine a spatio-temporal stochastic system with
multiple delays composed of coupled fiber ring lasers. A fiber ring
laser system without self feedback was studied in \cite{ShawSRR06},
and we extend the same model to include self feedback terms. 
In each ring laser, light circulates through a ring of optical
fiber, at least part of which is doped for stimulated emission. The
time for light to circulate through the ring is the cavity round-trip
time $\tau_{R}=202$ ns, and the delay time in the coupling and self 
feedback lines is a second delay $\tau_{d}=45$ ns.  Each
laser is characterized by a total population inversion $W(t)$  and an electric field $E(t)$.
The equations for the model dynamics of the $j^{th}$ laser are as
follows: \begin{eqnarray}
E_{j}(t) & = & R\exp\left[\Gamma(1-i\alpha_{j})W_{j}(t)+i\Delta\phi\right]E_{j}^{\text{fdb}}(t)\nonumber \\
 &  & +\xi_{j}(t)\label{Eequ}\\
\frac{dW_{j}}{dt} & = & q-1-W_{j}(t)\nonumber \\
 &  & -\left|E_{j}^{\text{fdb}}(t)\right|^{2}\left\{ \exp\left[2\Gamma W_{j}(t)\right]-1\right\} .\label{Wequ}\end{eqnarray}
 The electric field from earlier times which affects the field at
time $t$ is \begin{equation}
E_{j}^{\text{fdb}}(t)=E_{j}(t-\tau_{R})+\sum_{l\ne 
j}{\kappa_{l}E_{l}(t-\tau_{d})}+\kappa_{f}E_{j}(t-\tau_d).\label{Efdb}\end{equation}
$E_{j}(t)$ is the complex envelope of the electric field in laser
$j$, measured at a given reference point inside the cavity. $E_{j}^{\text{fbd}}(t)$
is a feedback term that includes optical feedback within laser $j$
and optical coupling with the other laser. Time is dimensionless.
Energy input is given by the pump parameter $q$. Each electric field
is perturbed by independent complex Gaussian noise sources, $\xi_{j}$,
with standard deviation $D$. We use a fixed input strength for all
coupling terms: $\kappa_{i}=\kappa_f=\kappa$ for all $i$.  (Values
of the parameters in the model as well as further computational details can be found
in \cite{ShawSRR06}.  The only difference in parameters was that the lasers are not detuned relative to each other in the current work.)

Because of the feedback term $E_{j}^{\text{fdb}}(t)$ in Eqs.~\ref{Eequ},
one can think of Eqs.~\ref{Eequ} as mapping the electric field on
the time interval $[t-\tau_{R},t]$ to the time interval $[t,t+\tau_{R}]$
in the absence of coupling ($\kappa=0$). Equivalently, because the
light is traveling around the cavity, Eqs.~\ref{Eequ} maps the electric
field at all points in the ring at time $t$ to the electric field
at all points in the ring at time $t+\tau_{R}$. We can thus construct
spatio-temporal plots for $E(t)$ or the intensity $I(t)=\left|E(t)\right|^{2}$
by unwrapping $E(t)$ into segments of length $\tau_{R}$.

\begin{figure}
\includegraphics[width=3.5in,height=3in,keepaspectratio]{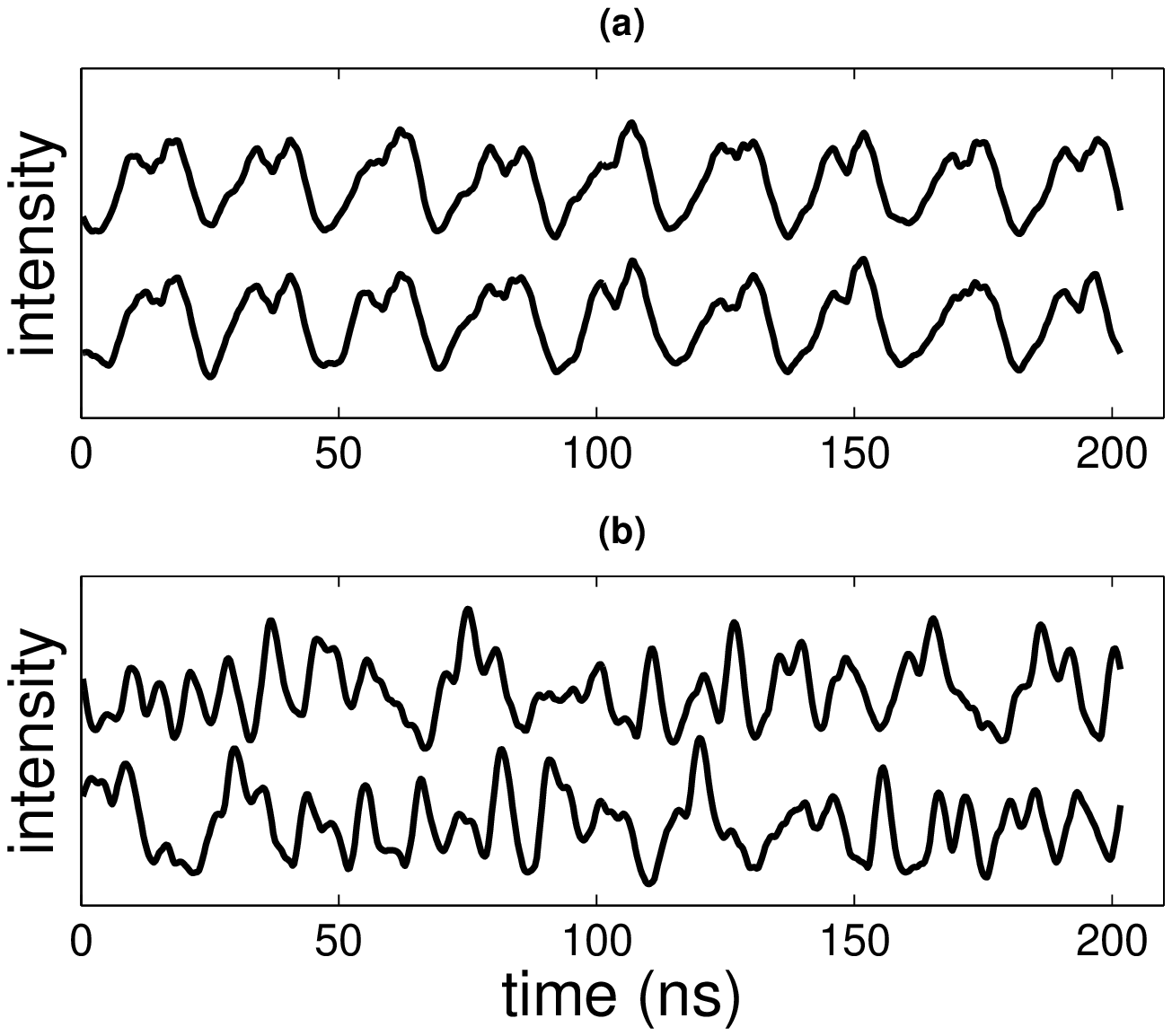}
\includegraphics[width=4.0in,height=5in,keepaspectratio]{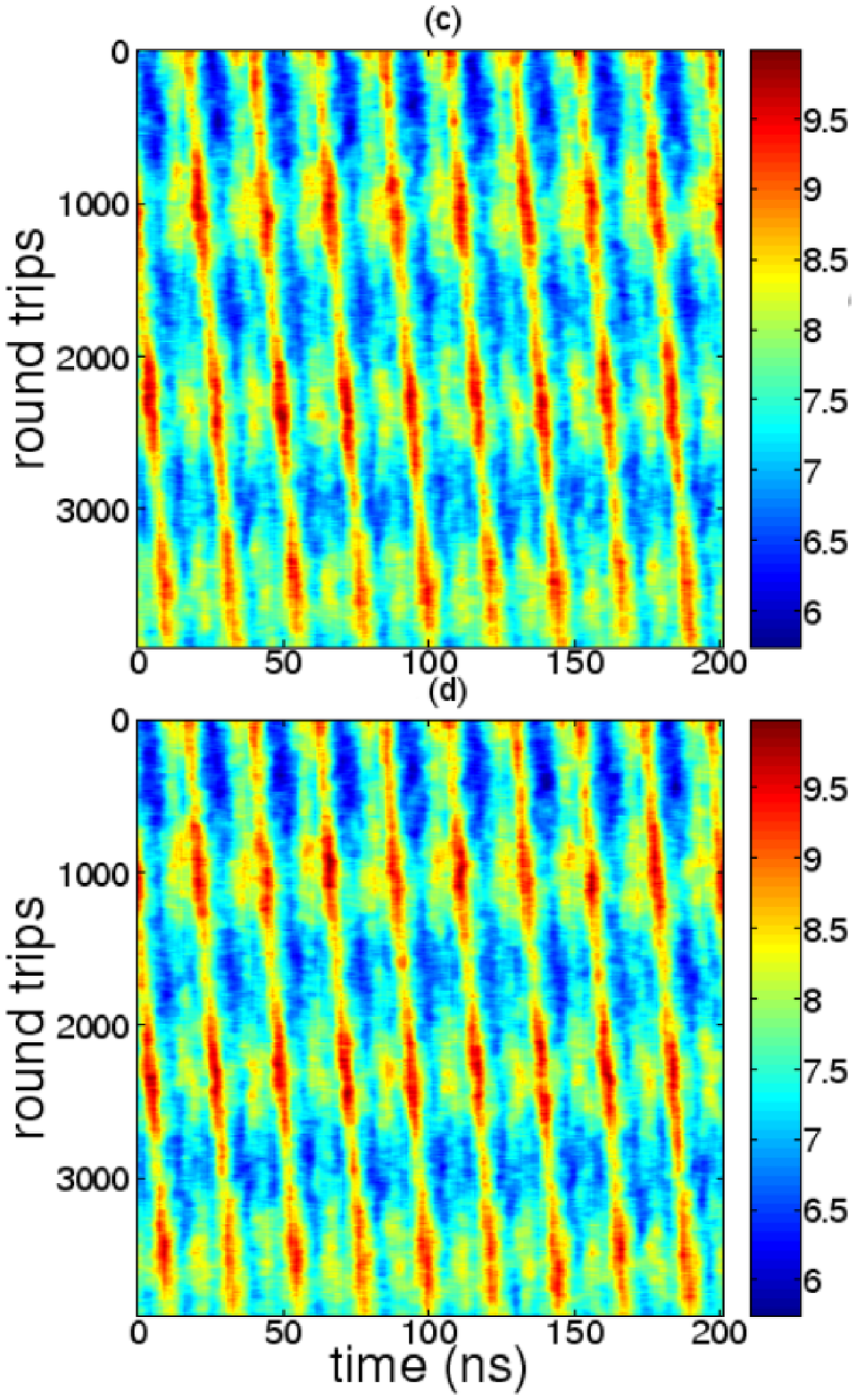}

\caption{(Color online) Intensity (arbitrary units) for two lasers coupled 
with $\kappa=0.009$.
The left panels are intensity vs.~time for laser 1 (bottom curve),
and for laser 2 (top curve): (a) With self feedback, (b) without
self feedback. Spatio-temporal plots corresponding to coupling with
self feedback for (c) laser 1 and (d) laser 2.}

\label{fig:2line_1rt_traces} 
\end{figure}

Figure \ref{fig:2line_1rt_traces} shows time traces of the $N=2$ lasers
for a single round trip for both the system with self feedback described
here and the system without self feedback ($\kappa_{f}=0$) \cite{note2}.
Isochronal synchrony can been seen when self feedback is included,
while in the absence of self feedback the lasers are delay synchronized.
The spatio-temporal plots in \ref{fig:2line_1rt_traces}(c) and (d) are nearly identical due to the isochronal synchrony. To quantify the synchrony, we align the time traces for the two lasers
with various time shifts between them.  In the absence of self feedback, the peak cross-correlation
occurs when the lasers are shifted relative to each other by the delay
time. The cross-correlation is low when the lasers are compared with
no time shift. In contrast, when self feedback is included, the lasers
achieve a high degree of correlation when compared isochronally. For the time traces shown in Fig.~\ref{fig:2line_1rt_traces}(a), the
peak cross-correlation of 0.9554 occurs when there is no time shift, although the cross-correlation when shifted by the delay time is near as high (0.9549).

\begin{figure}
\includegraphics[width=3.5in,keepaspectratio]{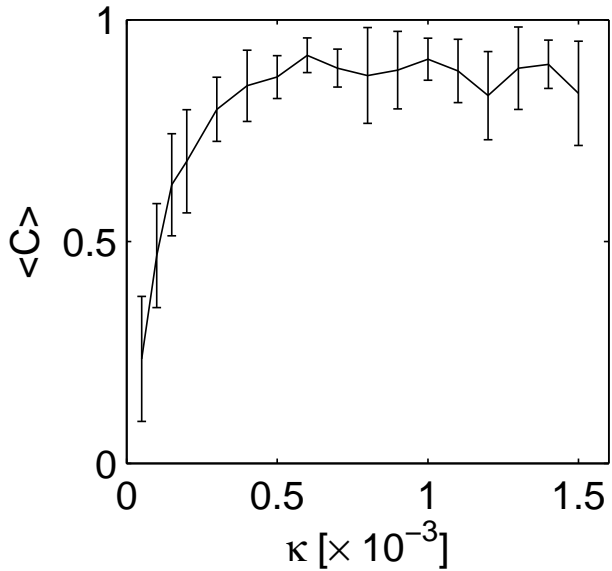}

\caption{Average cross-correlation vs.~coupling for two coupled lasers
with self feedback.}

\label{fig:2lineCvsk} 
\end{figure}

We have swept the coupling strength $\kappa$ for the system of two
lasers with self feedback and computed the average cross-correlation
when the lasers are compared isochronally. Figure \ref{fig:2lineCvsk}
 shows that the lasers are well synchronized for input strengths
as small as 0.1\%. \textbf{Isochronal synchronization can be produced when 
the 
lasers are detuned as in \cite{ShawSRR06}, but this requires stronger 
coupling and self feedback (not shown).}

For $N=3$ fiber ring lasers, we have done a similar
computation for cases with and without self feedback (not
shown). We found  that  when the
the lasers are coupled globally without self feedback, the isochronal state will still
synchronize. However, adding self feedback will cause the isochronal
state to stabilize at somewhat lower values of coupling. Further details
for this case are in \cite{LBIBS06}.

In summary, we have considered delay-coupled systems and, through the
addition of self feedback, obtained stable isochronal synchrony in
coupled semiconductor and fiber ring laser models. Model analysis for
incoherent pump coupled lasers reveals scaling of the
Lyapunov exponents transverse to the synchronized state, while
computations on systems of coupled fiber ring lasers show how self
feedback may cause the onset of synchrony in coupled spatio-temporal systems.  In the cases we have studied, we constructed small globally
coupled networks.  For the small clusters presented
here with delay, it is advantageous to add feedback loops, since
this was key to stabilizing the synchronous state. A question for future study
is how this method may 
be scaled up for larger networks.

The authors thank Raj Roy and Tony Franz for very stimulating disucssions and
comments. The authors also acknowledge the support of the Office of Naval Research.
LBS is currently a National Research Council Post Doctoral fellow.


\begin{thebibliography}{20}
\bibitem[{not({\natexlab{b}})}]{note2}
\bibinfo{note}{ The intensities displayed for the fiber laser are filtered with a 125 MHz low pass filter, to correspond with a typical setup for measuring intensities experimentally \cite{ShawSRR06}.  }
\end{thebibliography}

\end{document}